\documentclass[a4paper]{article}
\usepackage{graphicx}
\usepackage[margin=0.6in]{geometry}
\usepackage{amsmath}
\usepackage{amssymb}
\usepackage{cite}

\usepackage{multicol} %for two columns
\usepackage{caption}
\usepackage{hyperref}
\newcommand{\footremember}[2]{%
    \footnote{#2}
    \newcounter{#1}
    \setcounter{#1}{\value{footnote}}%
}
\newcommand{\footrecall}[1]{%
    \footnotemark[\value{#1}]%
} 

\usepackage[toc,page]{appendix}

\newcommand{\beginsupplement}{%
        \setcounter{table}{0}
        \renewcommand{\thetable}{S\arabic{table}}%
        \setcounter{figure}{0}
        \renewcommand{\thefigure}{S\arabic{figure}}%
        \setcounter{equation}{0}
        \renewcommand{\theequation}{\snum\arabic{equation}}
}
\usepackage{xr}
\newcommand{\snum}{S}

\title{Similarities between action potentials and acoustic pulses in a van der Waals fluid}

\author{Matan Mussel\footremember{TU-Dortmund}{Department of Physics, Technical University of Dortmund, 44227 Dortmund, Germany}\footremember{AugsburgUni}{Department of Physics, University of Augsburg, 86159 Augsburg, Germany}\footnote{Correspondence to: matan.mussel@physik.uni-augsburg.de}%
  \and Matthias F.~Schneider\footrecall{TU-Dortmund} %
  }

\begin{document}

\maketitle

\begin{abstract}
An action potential is typically described as a purely electrical change that propagates along the membrane of excitable cells. However, recent experiments have demonstrated that non-linear acoustic pulses that propagate along lipid interfaces and traverse the melting transition, share many similar properties with action potentials. Despite the striking experimental similarities, a comprehensive theoretical study of acoustic pulses in lipid systems is still lacking. Here we demonstrate that an idealized description of an interface near phase transition captures many properties of acoustic pulses in lipid monolayers, as well as action potentials in living cells. The possibility that action potentials may better be described as acoustic pulses in soft interfaces near phase transition is illustrated by the following similar properties: correspondence of time and velocity scales, qualitative pulse shape, sigmoidal response to stimulation amplitude (an `all-or-none' behavior), appearance in multiple observables (particularly, an adiabatic change of temperature), excitation by many types of stimulations, as well as annihilation upon collision. An implication of this work is that crucial functional information of the cell may be overlooked by focusing only on electrical measurements. 
\end{abstract}

{\bf Keywords:} action potential, acoustic pulse, lipid interface, phase transition, van der Waals fluid \hfill \break

%\section*{Significance Statement}
%Excitation processes in cells (action potentials) are principally associated with behavioral activities of many organisms. Therefore, the underlying mechanism of action potentials constitutes one of the fundamental aspects of biology. Several lines of evidence have suggested that the classical interpretation of these pulses as purely electrical was premature. Here it is demonstrated that non-linear acoustic pulses, which capture the phenomenology of action potentials, arise naturally from a minimum set of well accepted physical assumptions: mass, momentum and energy conservation laws near phase transition. The realization that action potentials are acoustic pulses, that carry electric changes as they travel, implies that significant information may be overlooked due to misunderstanding of excitable cells.

% ----------------- I n t r o d u c t i o n ----------------------------
\begin{multicols}{2}
\section{Introduction}
A large category of cells can generate a characteristic transient change in transmembrane voltage that propagates along the cell membrane in response to suitable stimuli. These include neurons, myocytes, epithelial cells, fibroblasts, glia cells, pancreas beta cells as well as non-specialized cells in corals, plants, fungi, Protozoa and possibly even bacteria\cite{Tasaki1982, Mackie2004, Parak1999, fields2008oligodendrocytes, Ashcroft1989, Leys2015, Beilby2007, slayman1976action, Wood1982, kralj2011electrical}. The phenomenon was first identified in a frog nerve by Emil du Bois-Reymond in 1843, who called them ``action currents'', later to be termed {\it action potentials}\cite{Tasaki1982}. Action potentials (APs) are principally associated with behavioral activities of many organisms. Therefore, an understanding of the mechanism of APs, as well as their actions and interactions, constitutes one of the fundamental aspects of biology. 

Classically APs are described as a purely electrical phenomenon, and the non-linear electric response is believed to be associate with specialized protein components\cite{Hodgkin1952, aidley1998physiology}. An unavoidable implication of this view is that the phenomenon should only exist in living cells, and could not be identified in non-living systems. However, pulses with similar properties have been observed in non-living soft systems such as lipids and gels\cite{Shrivastava2014, Przybylski1982}. In addition, many experimental facts are neither readily explained nor predicted by the electrical theory of APs. A partial list includes (1) non-electric aspects that co-propagate with the electric pulse\cite{Tasaki1999}; (2) the existence of APs in absence of sodium and potassium in the intra- and extracellular solutions\cite{Tasaki1966}, and in absence of ion-concentration gradient\cite{Terakawa1981}; and (3) ion-channel like current fluctuations in the absence of ion channel proteins\cite{Wunderlich2009}. Therefore, the classical picture not only is not derived from fundamental physical principles, it also does not satisfactory describe the observed phenomenology. Alternative approaches, that treat the AP on a more physicochemical basis, have been proposed\cite{Tasaki1999, Heimburg2005, Kaufmann1989, ling1992revolution}. A common thread of these approaches is that electricity is merely one aspect of the pulse, and it is, therefore, very likely that valuable information is overlooked due to misunderstanding of its mechanism.

One conjecture, which is the focus of this manuscript, is that APs are acoustic pulses that propagate along the lipid interface and cross the phase-transition from the so called {\it liquid-expanded} to the {\it liquid-condensed} phase \cite{Kaufmann1989, Heimburg2005}. Indeed, experimental observations in lipid monolayers have demonstrated that solitary acoustic pulses can be excited, and that they share many properties with APs. A comparison of key experimental observations between APs and acoustic pulses in lipid interfaces is as follows. (1) APs are not purely electrical waves. Rather, they were identified in multiple observables -- electrical, magnetic, mechanical, optical, as well as thermal\cite{Hodgkin1945, Tasaki1999, Wikswo1980}. Acoustic pulses in lipids were also identified in multiple observables (density, pressure, electrical, optical and pH)\cite{Steppich2010, Griesbauer2012, Shrivastava2014, Shrivastava2015, Fichtl2016}. Particularly, the electric potential difference between the liquid-expanded and the liquid-condensed phase is $\sim$100 mV, similar to the amplitude of APs\cite{Steppich2010}. (2) The production and absorption of heat during an AP is adiabatic\cite{Ritchie1985}, and an acoustic description is the natural physical approach for an adiabatic propagating pulse. (3) The time and velocity scales of APs vary by 5--6 orders of magnitude between cells, $\sim 10^{-3}-10^1$ s and $\sim 10^{-3}-10^2$ m/s, respectively\cite{Hodgkin1952, Leys2015, Beilby2007}. These values are comparable to sound pulses in lipid systems that traverse the phase transition ($\sim10^{-3}-10^{-2}$ s and $\sim$0.1--1 m/s, respectively)\cite{Shrivastava2014, Shrivastava2015}. (4) Excitation of an AP is obtained by many types of stimulations: electrical, mechanical (pressure, touch, ultrasound), optical (by shining light), as well as a change of temperature (heating or cooling)\cite{Heimburg2007, Tasaki1982}. Acoustic pulses in lipids can also be excited by various types of stimulations (mechanical, electrical and chemical (acid as well as solvent))\cite{Griesbauer2009, Griesbauer2012a, Shrivastava2014, Fichtl2016}. (5) Even the less intuitive properties of APs, namely threshold behavior (a sigmoidal response, the so called {\it all-or-none})\cite{Hodgkin1952a} and annihilation upon collision\cite{Tasaki1949, Follmann2015}, can be demonstrated as acoustic behavior near phase transition. Specifically, stimulation was shown to prompt a sigmoidal response of density pulses in lipid interfaces near phase transition\cite{Shrivastava2015}, and these pulses were demonstrated to annihilate upon collision\cite{Shrivastava2017}.

Despite the striking experimental similarities, a comprehensive theoretical study of acoustic pulses that traverse the phase transition in lipid systems is still lacking. Theoretical modeling was initiated by Heimburg and Jackson, and was focused on solitonic solutions in a small-amplitude analysis of an acoustic model of the membrane interface\cite{Heimburg2005}. Two main criticisms of the model have been an overestimation of the propagation velocity and the lack of annihilation of pulses. Kappler et al.~later demonstrated that the viscous coupling between the interface and a bulk fluid attenuates the propagation velocity of interfacial longitudonal waves, and results in a corrected velocity that agrees with experimental evidence\cite{Kappler2017}. Still, some observations are not captured by these previous works, including the saturation of pulse amplitude at large stimulation amplitudes, annihilation of pulses upon collision, as well as a thorough analysis of different aspects of the pulses under adiabatic conditions (e.g., density, pressure, temperature and electrical properties). 

Herein, we demonstrate that non-linear acoustic pulses, which annihilate upon collision, arise naturally from a minimum set of well accepted physical assumptions: mass, momentum and energy conservation laws near a phase transition. Our description relies on relatively few and physically measurable parameters and variables. Particularly, no assumptions about the molecular constituents and structure, nor any parameter fit, are required. In spite of its minimalism, the model provides a rich behavior of non-linear acoustic pulses that correspond to experimental results in lipid monolayers as well as to APs in living cells. 

% ----------------- M o d e l    D e s c r i p t i o n  ----------------------------
\section{Model Description}
Our purpose is to investigate the properties of isentropic traveling waves which exist near a phase transition. We therefore consider an idealized ansatz of a one-dimensional compressible fluid with a van der Waals constitutive equation, following the works of Felderhof, Slemrod and Grinfeld\cite{slemrod1984dynamic, grinfeld1989nonisothermal, felderhof1970dynamics}. While classically a van der Waals phase transition is between a gas and a liquid, the familiar pressure-versus-volume curves (isotherms) qualitatively resemble the melting transition between the liquid-expanded and the liquid-condensed state in lipids\cite{Albrecht1978}. The conservation laws of mass, momentum and energy are written in the Lagrange frame,
\begin{equation}\label{eq:consLaws}
\begin{aligned}
\partial_t w&=\bar{w}\partial_h v, \\
\partial_t v&=\bar{w}\partial_h (\tau_1+\tau_2 ), \\
\partial_t E&=\bar{w}\left[\partial_h (\tau_1 v)+k\partial_h^2 \theta\right].
\end{aligned}
\end{equation}
Here $w, v$ and $E$ denote the specific volume, velocity and specific total energy of the fluid, respectively. The fluid density is inversely proportional to the specific volume, $\rho=w^{-1}$. The spatial coordinate $h$ (in the Lagrange frame) is related to the x-axis (in the Euler frame) by the cumulative mass of fluid particles\cite{courant1948supersonic}
\begin{equation}
x=\frac{1}{\bar{w}}\int w dh,
\end{equation}
with $\bar{w}$ a normalization factor that defines the scale of $h$. In addition, $\tau_1$ and $\tau_2$ represent the stresses in the fluid (see below), $k$ is the coefficient of thermal conductivity and $\theta$ is the fluid temperature.

To allow for a continuous transition between the two phases, we follow van der Waals' hypothesis that the energy of a system depends on the gradient of density (the so called {\it capillarity} term)\cite{Gorban2016, van20thermodynamic}. A similar definition appears in the Ginzburg-Landau mean field theory of phase transition as well as the Cahn-Hillard model of coexistence of two phases\cite{landau1980statistical, cahn1958free}. The capillarity coefficient is also known as the {\it gradient-energy coefficient}. The corresponding stress correction to the dynamics of fluids was developed by Korteweg\cite{Gorban2016, korteweg1901forme}. Therefore, the expression for stress includes a capillarity term in addition to pressure and viscosity. It is defined as
\begin{equation}
\begin{aligned}
\tau_1&=-p+\zeta\partial_h v, \\
\tau_2&=-C\partial_h^2 w,
\end{aligned}
\end{equation}
with $p$ the fluid pressure, $\zeta$ the dilatational viscosity and $C$ the capillarity coefficient, treated here as constant for simplicity. 
\begin{figure*}[htb]
\centering
\includegraphics[width=0.6\linewidth]{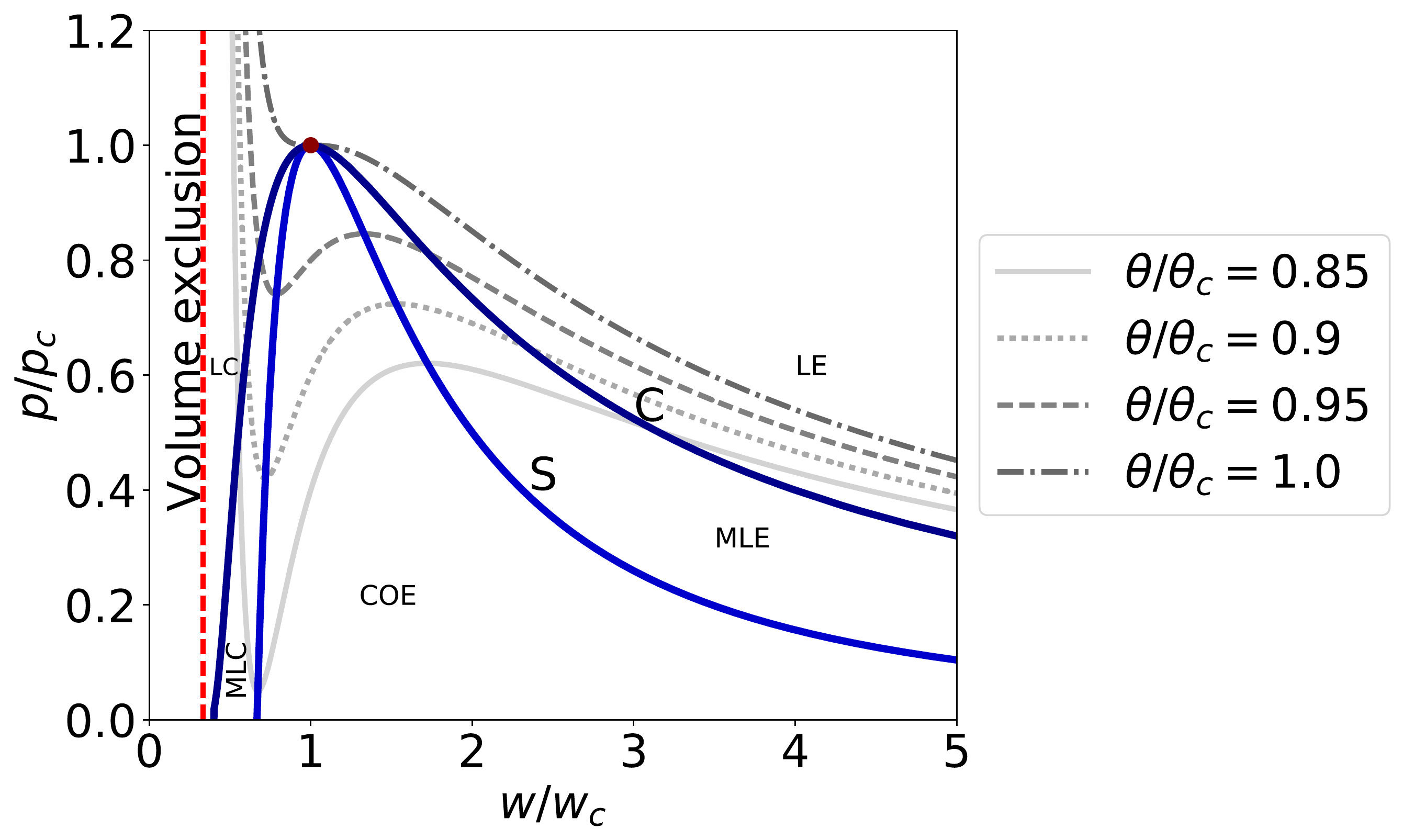}
 \caption{Schematic phase diagram in the w--p plane of the van der Waals model. Four different isotherms, along with the Spinodal (S) and coexistence (C) curves are plotted. The critical point is denoted by a filled circle. LE: liquid-expanded phase, MLE: metastable liquid-expanded phase, COE: coexistence phase, MLC: metastable liquid-condensed phase, and LC: liquid-condensed phase.}
 \label{Fig-01}
\end{figure*}

The system of equations is completed with the van der Waals constitutive equations, that qualitatively resemble the transition between a liquid-expanded and a liquid-condensed state in lipids (Fig.~\ref{Fig-01} and Ref.~\cite{Albrecht1978})
\begin{equation}\label{eq:vdW}
\begin{aligned}
p&=\frac{k_B \theta}{m w - b} - \frac{a}{m^2 w^2},\\
E&=c_v \theta - \frac{a}{m^2 w} + \frac{v^2}{2},
\end{aligned}
\end{equation}
with $k_B$ the Boltzmann constant, $m$ the mass of a fluid particle, $a$ the average attraction between particles, $b$ the volume exclusion by a fluid particle, and $c_v$ the specific heat capacity\cite{johnston2014thermodynamic}. The phase space contains an unstable region, $\partial p/\partial w >0$. This region is bounded by the spinodal curve (marked by S in Fig.~\ref{Fig-01}), which satisfies $\partial p/\partial w =0$. A slightly larger area, governed by the Maxwell equal area rule, describes the region where a coexistence of phases has a lower free energy as compared to a single phase solution (marked by C in Fig.~\ref{Fig-01}). Both curves meet at the critical point $(w_c, p_c, \theta_c)$, where the distinction between the two phases disappears. The critical point can be expressed by three parameters: $m, a$ and $b$ (Eq.~(S1) in the Supplemental Materials). A detailed list of the 6 variables and 7 parameters of the model is provided in the Supplemental Materials (Tables S1--S2). 

The critical point and the fluid viscosity were used to define proper scales (time, length and velocity, respectively)
\begin{equation}\label{eq:scales}
T\equiv\frac{\zeta}{p_c} ,\quad L\equiv\zeta\sqrt{\frac{w_c}{p_c}} , \quad U\equiv\frac{L}{T}=\sqrt{w_c p_c}.
\end{equation}
These scales were used to derive a dimensionless form of the model equations (Supplemental Materials, Eqs.~(S3)--(S5)), that depends on only three parameters: the (dimensionless) heat capacity, thermal conductivity and capillarity coefficient
\begin{equation}
\tilde{c}_v=\frac{c_v \theta_c}{p_c w_c}, \quad \tilde{k}=\frac{k\theta_c}{p_c w_c \zeta},\quad \tilde{C}=\frac{C}{\zeta^2} .
\end{equation}
Dimensionless variables and parameters are hereafter marked with tilde (e.g., $\tilde{x}=x/L$).
     
\begin{figure*}[htb]
\centering
\includegraphics[width=0.9\linewidth]{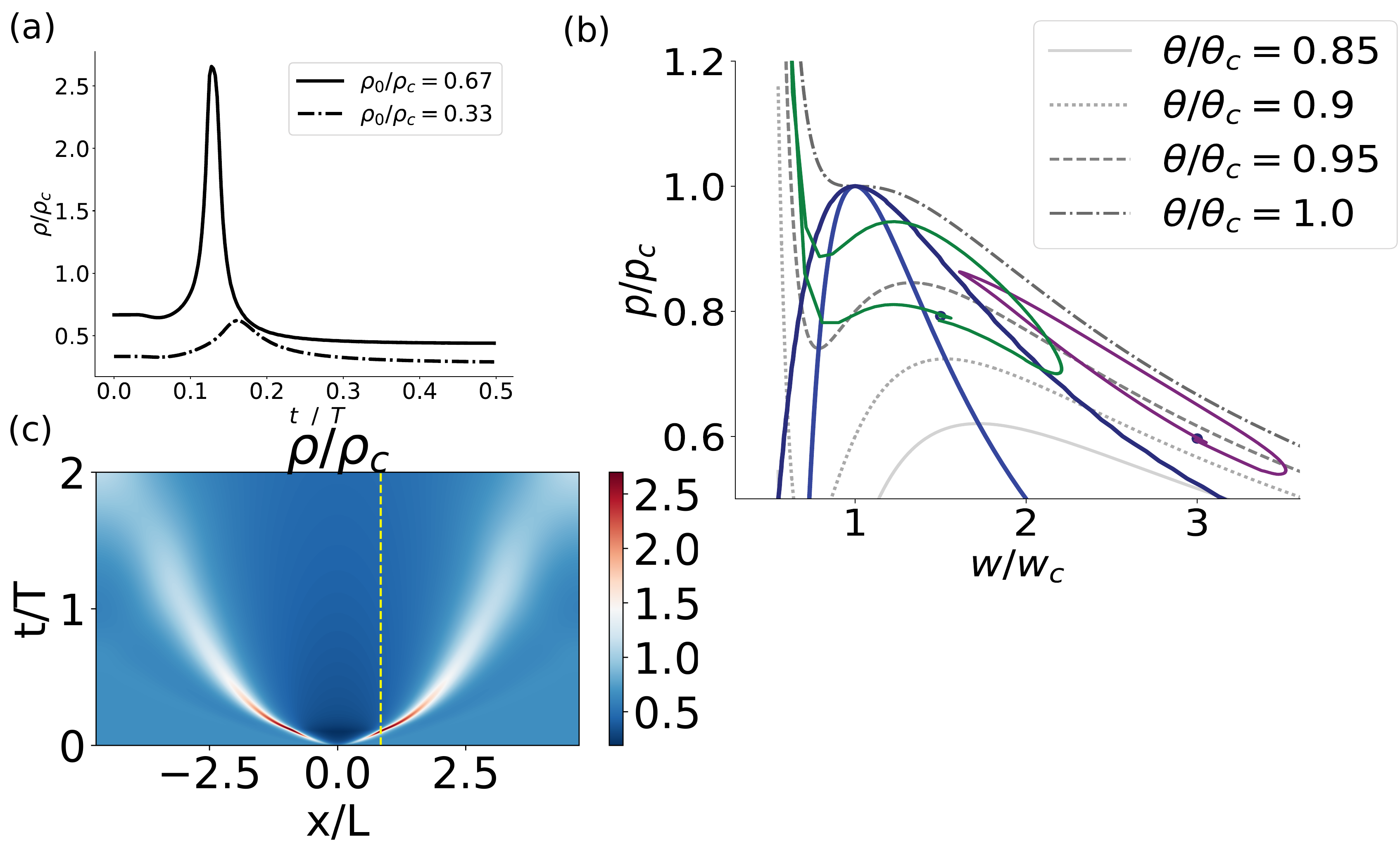}
 \caption{(a) Density pulse as a function of time, as measured at distance $x/L=1$ from the excitation point, with an initial density of $\tilde{\rho}_0$=0.67 (solid line) and 0.33 (dotted-dashed line). (b) A projection of phase space into the w--p plane (w=$\rho^{-1}$). Several isotherms are plotted for reference (shades of grey lines) as well as the coexistence and spinodal curves (dark blue and blue lines, respectively). The trajectory of the two pulses shown in (a) is plotted in green and purple respectively. (c) Density field of the entire fluid as a function of space (x-axis) and time (y-axis) with initial density $\tilde{\rho}_0$=0.67. Dashed yellow line marks the solid line solution depicted in (a). Parameters of the model were $(\tilde{c}_v, \tilde{k}, \tilde{C})$=(600, 100, 1), additional initial conditions were $(\tilde{v}_0, \tilde{\theta}_0)$=(0, 0.93), and excitation parameters were $(\tilde{x}_0, \tilde{t}_0, \tilde{p}_{exc}, \lambda)$=(0, 0.1, 150, 0.088). Numerical calculation was conducted with 4096 grid points, x-domain $[-3\pi/2, 3\pi/2]$ and $dt=5\cdot 10^{-4}$.}
 \label{Fig-02}
 \end{figure*}
     
A {\it complete} set of all of the material constants is not known for any {\it single} lipid system, nor for a composite soft interface, for example in biological cells. Therefore, typical values were estimated from experiments with DPPC lipids. The critical point of a  two-dimensional DPPC monolayer was identified from isothermal state diagrams $(w_c, p_c, \theta_c)\cong(5\cdot 10^5~m^2/kg, 3\cdot10^{-2}~Pa\cdot m, 315~K)$\cite{Albrecht1978, Heimburg2007, Steppich2010, Shrivastava2014}. The order of magnitude of the membrane viscosity was estimated from measurements of shear viscosity, $\zeta\sim10^{-3}~Pa\cdot s\cdot m$\cite{espinosa2011shear}. The order of magnitude of the heat capacity was evaluated from experiments at constant pressure, $c_v\sim c_p\sim10^3-10^4~J/kg\cdot K$\cite{Steppich2010, Heimburg2005}. Unfortunately, no direct measurements of the thermal conductivity and capillarity coefficient of lipid monolayers exist. Thermal conductivity was approximated from the evaluation of heat conduction in interfacial water in a lipid system, $k\sim5~J/s\cdot K$\cite{clegg1991biochemistry}. The capillarity coefficient was estimated from measurements of line tension at the phase boundary, $C\sim10^{-27}-10^{-24}~kg^2/s^2$ (Supplemental Materials and Ref.~\cite{Benvegnu1992}). For typical values of a lipid system, the proper scales are T$\approx$30 ms, L$\approx$4 m and U$\approx$120 m/s, and the dimensionless parameters are $\tilde{c}_v\sim 10^1-10^3$, $\tilde{k}\sim 10^2$, and $\tilde{C}\sim 10^{-21}-10^{-18}$. Although the capillarity coefficient is insignificantly small, a non-negligible value was used for numerical reasons ($\tilde{C}\sim 1$) -- to avoid sharp gradients in the density field. The use of a non-negligible capillarity coefficient did not show a major effect on our results (Fig.~S1). 

The system of equations (\ref{eq:consLaws})--(\ref{eq:vdW}) was numerically solved with the Dedalus open-source code\cite{Burns2017}, which is based on a pseudo-spectral method. The model was solved using periodic boundary conditions, and with homogeneous initial conditions, $(w_0, p_0, \theta_0)$ in the liquid-expanded phase. Excitation of a pulse was obtained by `injecting' a localized stress (around $h_0$) with an amplitude $p_{exc}$ for a brief time ($t_0$) into the momentum flux; i.e., adding the following term into the right-hand-side of the middle Eq.~\ref{eq:consLaws}
\begin{equation}\label{eq:excitationCurrent}
\bar{w}\partial_h \left(p_{exc} \Theta(t_0-t ) e^{-\frac{(h-h_0)^2}{2\lambda^2}} \right).
\end{equation}
Here, $\Theta$ is the Heaviside function, and $\lambda$ is the width of excitation. 

\begin{figure*}[htb]
\centering
\includegraphics[width=0.6\linewidth]{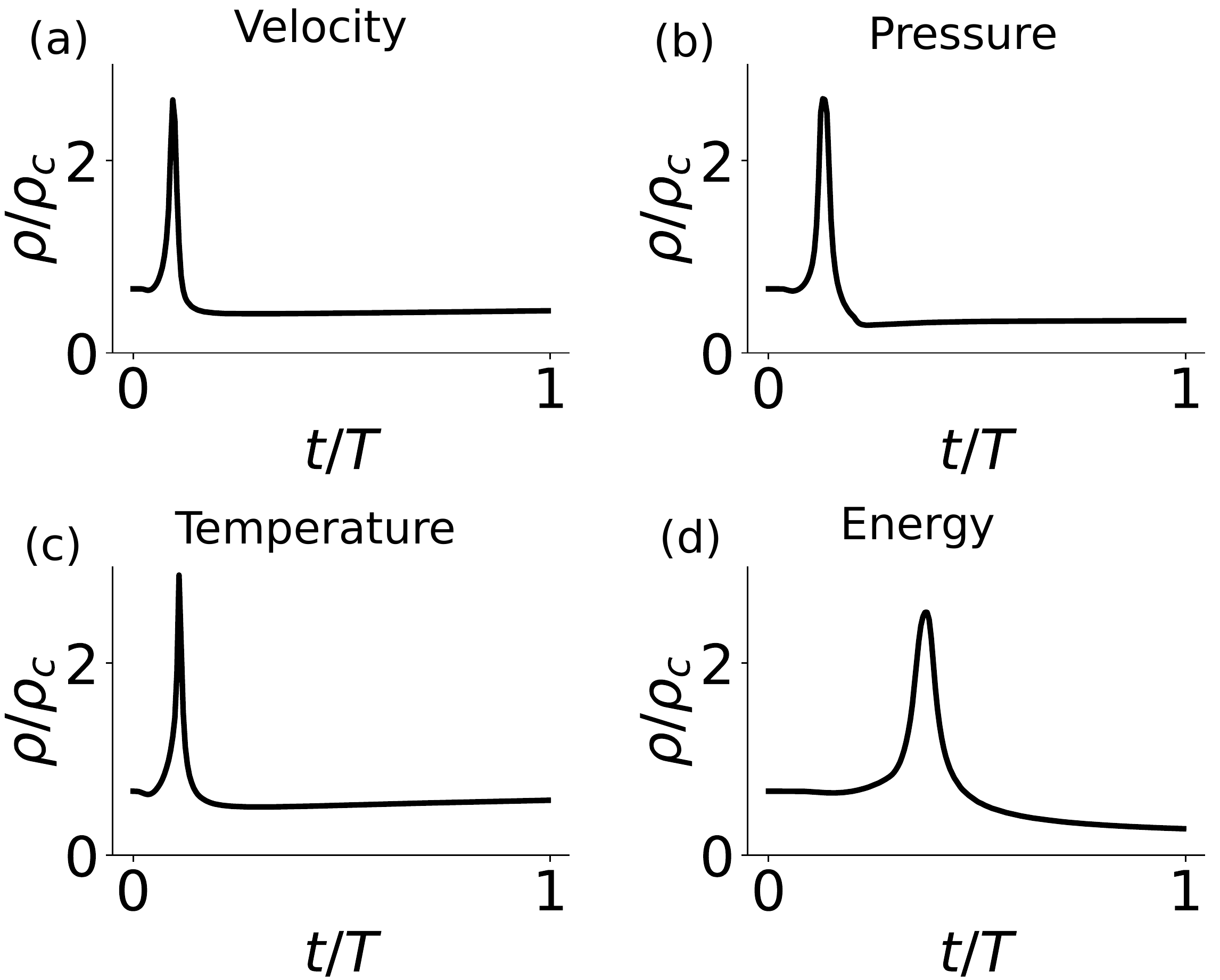}
 \caption{Excitation by a local injection of (a) velocity (middle of Eq.~(\ref{eq:consLaws})), (b) pressure (upper Eq.~(\ref{eq:vdW})), (c) temperature (lower Eq.~(\ref{eq:vdW})), and (d) energy (lower Eq.~(\ref{eq:consLaws})). Excitation parameters were $(\tilde{A}, t_0)=(150, 0.1), (150, 0.2), (1500, 0.1)$ and $(10^4, 0.2)$, respectively. $A$ is the normalized amplitude of excitation ($p_{exc}/p_c, p_{exc}/p_c, \theta_{exc}/\theta_c$ and $E_{exc}/U^2$, respectively). Density as a function of time was plotted at x/L=0.8, 1.0, 0.8 and 1.4, respectively.}
 \label{Fig-03}
 \end{figure*} 

% ----------------- R e s u l t s  ----------------------------
\section{Results}
We now turn to analyze sound pulses that traverse the phase transition. Upon excitation, distinct pulses were obtained when the initial state of the fluid was near phase transition. Figure \ref{Fig-02}a (solid line) depicts the shape of a density pulse as measured at distance $x/L=1$ from the excitation point. The shape of the density pulse was qualitatively very similar to experimental measurements in lipid monolayers\cite{Shrivastava2014}, as well as to the voltage signal of an AP\cite{Hodgkin1952a}. Because this issue is of particular relevance to the open debate regarding the physical nature of cellular excitability, we further discuss the relation between the interface density and a transmembrane voltage measurement in the Discussion section.  In contrast, pulse amplitude was much lower when the initial state was away from the phase transition (Fig.~\ref{Fig-02}a, dotted-dashed line). A projection of the closed trajectory in the w--p plane is shown in Fig.~\ref{Fig-02}b, for the two pulses depicted in Fig.~\ref{Fig-02}a (green and purple curves, respectively). If a trajectory does not cross the phase boundary (dotted-dashed line in Fig.~\ref{Fig-02}a and purple curve in Fig.~\ref{Fig-02}b) the pulse undergoes only a little amplification in density ($\tilde{\rho}\lesssim$1). The change in density is accompanied by a parallel increase in pressure ($\tilde{p}\lesssim$1) and temperature. Subsequently, the local state decreases in all three fields (density, pressure and temperature) into a rarefaction state before relaxing back into the initial state. In contrast, once a pulse traverses the phase transition region, the density and pressure evolve non-linearly (solid line in Fig.~\ref{Fig-02}a and green curve in Fig.~\ref{Fig-02}b). At first, a significant increase in density is obtained ($\tilde{\rho}\gtrsim$1), with almost no change in pressure. Subsequently, as the local state approaches the volume exclusion region, the pressure sharply increases ($\tilde{p}\gtrsim$1), with only a slight increase in density. A parallel small increase in temperature also occurs. Relaxation follows in a reverse order -- a sharp decrease in pressure is followed by a sharp decrease in density into a rarefaction that relaxes back into the initial state. 
 
The solution of the density field across the entire fluid is plotted in Figure \ref{Fig-02}c. Following an excitation around x=0 that lasted 0.1T, two localized pulses were generated, propagating in opposite directions. Their length, time and velocity scales were $\approx$0.5L, $\approx$0.1T, and $\approx$5U, respectively. An important observation is that these pulses did not maintain a constant velocity and shape, and eventually dispersed; i.e., these solutions are neither solitons, nor homoclinic orbits. Nonetheless, the pulses had a distinct shape for a distance of 3--5 times the pulse width. Variation of the duration or width of the stimulation ($t_0$ and $\lambda$ in Eq.~(\ref{eq:excitationCurrent}), respectively) did not have much influence on the time and length scales of the pulse (Fig.~S2). However, upon increasing the duration of the stimulation ($t_0$), the pulses maintained a distinct shape for a longer distance (more than 10 times the pulse width, Fig.~S2b). The dashed yellow line in Fig.~\ref{Fig-02}c marks the solution that was plotted in Fig.~\ref{Fig-02}a (solid line). 

Stimulation by different model variables (velocity, pressure, temperature or energy) was obtained by adding the excitation term ($\tilde{A} \Theta(t_0-t ) e^{-\frac{(h-h_0)^2}{2\lambda^2}}$)
%(Eq.~(\ref{eq:excitationCurrent})) 
to other model equations (Eqs.~(\ref{eq:consLaws}) or (\ref{eq:vdW})). Figure \ref{Fig-03} shows the density pulses that were generated by different `types' of stimulations. The resulting pulses were qualitatively similar. 
 
\begin{figure*}[htb]
\centering
\includegraphics[width=0.9\linewidth]{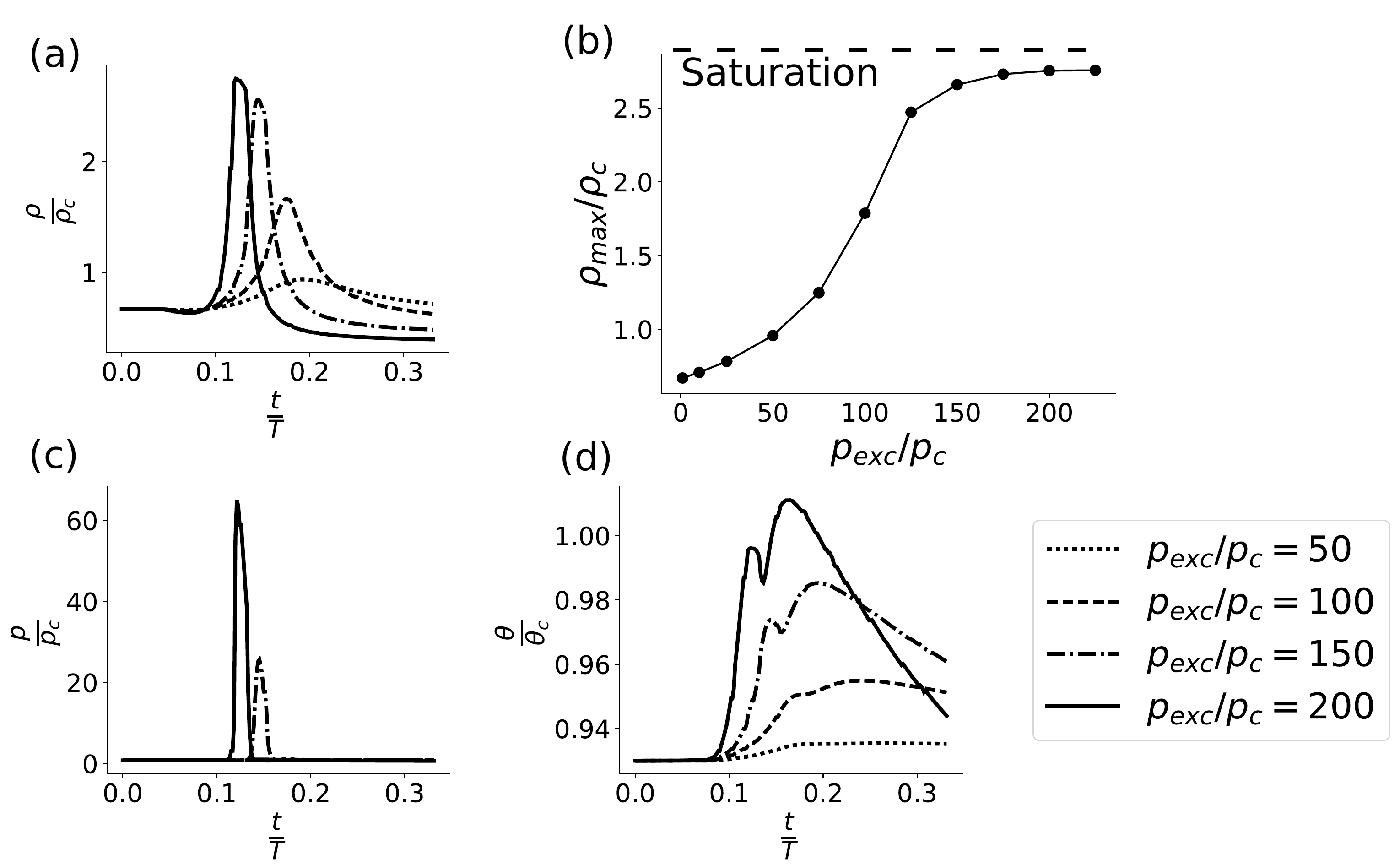}
 \caption{(a) Pulse shape at four amplitudes of the excitation, as reflected by the change of density. (b) Non-linear response of the amplitude of the density pulse ($\rho_{max}$) to stimulation amplitude ($p_{exc}$). (c) Pressure and (d) temperature aspects that co-appear with the density pulse. The first two pressure curves are indistinguishable, as the state did not reach the liquid-condensed phase. Initial density was $\tilde{\rho}_0$=0.67, and pulse was measured at distance $x/L$=1.1 from the excitation point. Other parameters appear in the caption of Fig.~\ref{Fig-02}.}
 \label{Fig-04}
 \end{figure*}

By increasing the amplitude of excitation, at a given initial state, a non-linear (sigmoidal) response of the density pulse was obtained (Figs.~\ref{Fig-04}a,b). At low amplitudes of excitation the pulse response was qualitatively similar to previous theoretical results obtained in a small-amplitude analysis\cite{Kappler2017}. However, at larger amplitudes of excitation the amplitude of the density pulse saturated at $\rho$=3$\rho_c$. The saturation of density is a direct result of the exclusion of volume given by the van der Waals equation (upper Eq.~(\ref{eq:vdW}) or (S5)). The response resembles experimental observations in lipid monolayers\cite{Shrivastava2015} as well as voltage measurements of APs in living cells\cite{Hodgkin1952a}.
 
Figure \ref{Fig-04}c shows the simultaneous pressure aspect of the pulse. Pressure increases significantly when the system is driven into the liquid-condensed phase. In addition, during the adiabatic compression and decompression, the temperature increases and subsequently decreases (Fig.~\ref{Fig-04}d). Not surprisingly, the temporal width of the temperature aspect crucially depends on the thermal conductivity (data not shown).

\begin{figure*}[htb]
\centering
\includegraphics[width=0.9\linewidth]{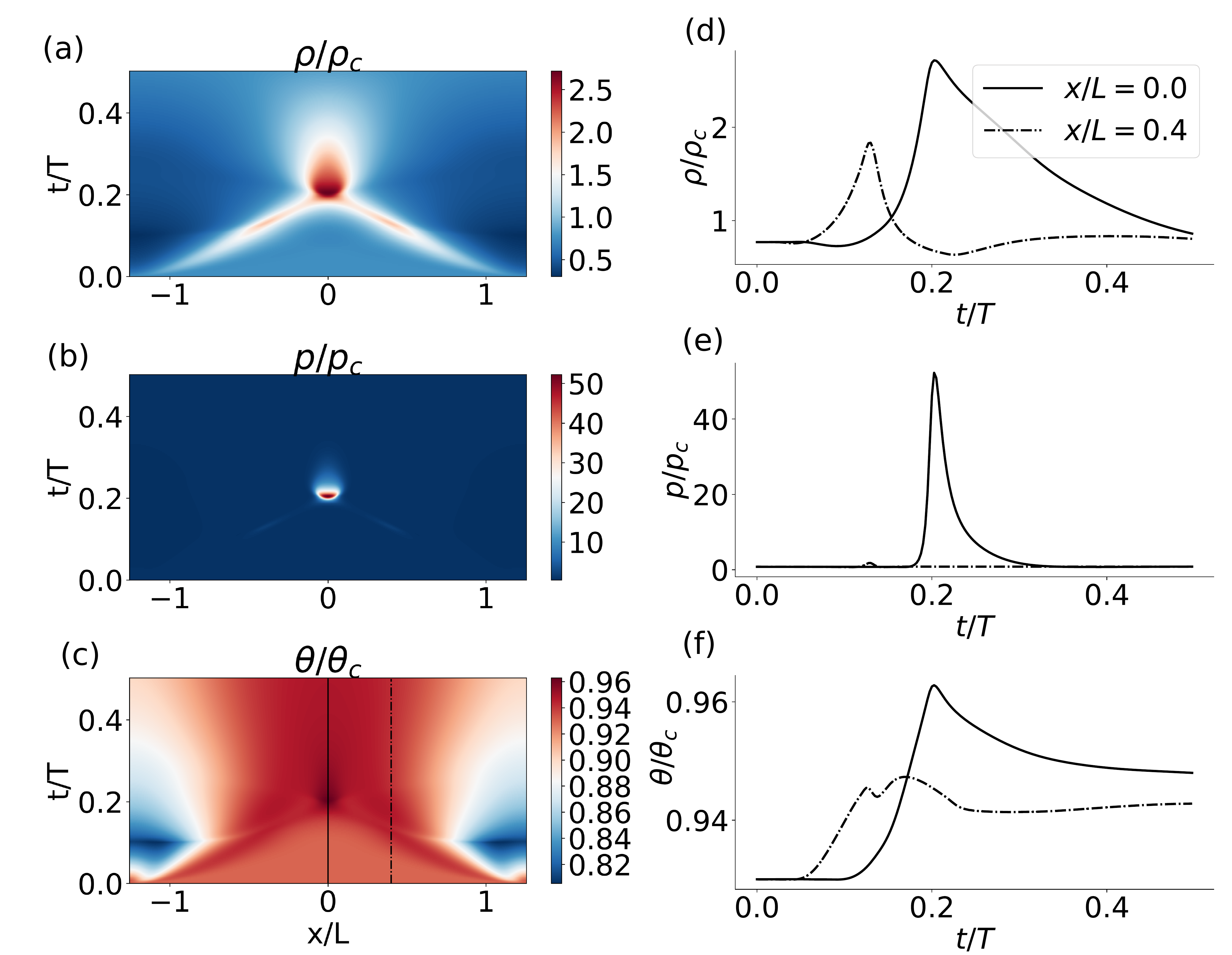}
 \caption{Collision between two pulses as appeared in the (a) density, (b) pressure and (c) temperature fields. x- and y-axis represent space and time, respectively. A local description of the (d) density, (e) pressure and (f) temperature is plotted as a function of time at the collision point (x/L=0, solid line) and at some distance away from it (x/L=0.4, dashed-dotted line). The location of these solutions in space is marked in (c) by solid and dashed-dotted line, respectively. Fluid initial density was $\tilde{\rho}_0$=0.77. Excitation was conducted at $\tilde{x}_0$=$\pm$0.4$\pi$. Numerical calculation was performed with 8192 grid points, and the x-domain was $[-0.4\pi, 0.4\pi]$. Other parameters are similar to those given in caption of Fig.~\ref{Fig-02}.}
 \label{Fig-05}
 \end{figure*}

We proceed now to investigate the question of collision between longitudonal pulses. While linear (small amplitude) pulses generally superimpose, these non-linear waves displayed a rich behavior of interaction, annihilation and in some cases even repulsion. The type of interaction between pulses strongly depends on the value of the heat capacity and thermal conductivity of the fluid. Here we provide an example of annihilation of two pulses that were excited in a small sized domain. Figure \ref{Fig-05} and Movie S1 show the dynamics of the density, pressure and temperature fields before, during and after a collision. The propagation of the pulses is evident in all three fields, and the collision is characterized by a localized increase in amplitude, most dramatically observed in the pressure field.

% ----------------- D i s c u s s i o n  ---------------------------- 
 \section{Discussion}
The lipid-based interface is a ubiquitous component of biological cells, and plays a critical role in many cellular functions. Particularly, the discovery of solitary longitudinal pulses that travel along lipid interfaces have stimulated a discussion on the functional role of acoustics in biological systems\cite{Heimburg2005, Kaufmann1989, Heimburg2007, Griesbauer2009, Griesbauer2012, Shrivastava2014, Shrivastava2015, Fichtl2016}. In this paper we continue this line of research and demonstrate that an idealized fluid model near phase transition captures many properties of experimentally measured acoustic pulses in lipid monolayers as well as APs in living cells.
 
\paragraph{Time, length and velocity scales of density pulses in lipid interfaces.} The critical point and the fluid viscosity specify proper scales of time, length and velocity (Eq.~(\ref{eq:scales})). Specifically, density pulses that traverse the phase transition region scale as $\approx$0.1T, $\approx$0.5L and $\approx$5U, respectively (Fig.~\ref{Fig-02}). For typical lipid values these scales correspond to $\approx$3 ms, $\approx$2 m, and $\approx$600 m/s. In comparison, experimentally measured pulses in lipid monolayers scale as $\approx$5--10 ms, $\approx$1--10 mm and $\approx$0.1--1 m/s\cite{Shrivastava2014, Shrivastava2015}, respectively. While the theoretical time scale agrees with measurements, the length and velocity scales are overestimated by 2--3 orders of magnitude (the same overestimation of velocity was previously obtained by others\cite{Heimburg2005, Griesbauer2009}). The discrepancy in velocity and length scales results from our attempt to keep the model simple, and would disappear by extending the model to include the non-negligible coupling to the bulk. It was previously demonstrated that the viscous coupling between a compressible interface and the bulk attenuates the velocity and length of acoustic pulses by a factor of $\sim \sqrt{1+\sqrt{\rho_b\eta_b t_p}/\rho_i}$\cite{Kappler2017}. Here, $\rho_b$ is the bulk density, $\eta_b$ the bulk viscosity, $t_p$ the pulse duration, and $\rho_i$ the density of the interface. For a typical lipid interface coupled to bulk water, this results in attenuation by 2--3 orders of magnitudes, and agrees with experimental observation\cite{Griesbauer2009, Kappler2017}. Furthermore, these scales also agree with experimental measurements of APs in living cells ($\sim 10^{-3}-10^1$ s and $\sim 10^{-3}-10^2$ m/s, respectively\cite{Hodgkin1952, Leys2015, Beilby2007}). The variation of 5--6 orders of magnitudes in time and velocity of APs may be related to differences in the macroscopic properties of the cell surface (critical temperature, surface viscosity, relative state as compared to the critical point), as well as the bulk (density and viscosity).

\paragraph{Dispersion of shape and propagation velocity.} According to the classical electrical description, APs are pulses that maintain a stable shape and a constant propagation velocity along an `infinitely' long cell (a homoclinic orbit)\cite{Hodgkin1952}. Similar characteristics were also suggested by the acoustic soliton model\cite{Heimburg2005}. Surprisingly, a validation of this hypothesis, by physiological measurements of velocity and shape at more than two sites, was hardly investigated, presumably due to the small cellular size. However, recent experiments have showed, using a multi-electrode array, that the propagation velocity as well as the shape of an AP are not constants during propagation. Rather, a clear trend of decrease in velocity and amplitude, as well as an increase in width, was observed\cite{Bakkum2013, patolsky2006detection}. These findings are in accord with our results, that do not preserve a constant velocity and shape, and eventually disperse. We have, nonetheless, demonstrated that distinct pulses were maintained at distances up to 1--10 times the length of a pulse (Figs.~\ref{Fig-02}c and S2b). Interestingly, the length of many excitable cells is of similar order. For example, in {\it Loligo Pealei} squids the length of a giant axon is 3--5 times the length of an AP (the axonal length is 10--20 cm, and the estimated length of an AP is $\approx$4 cm (the pulse duration is $\approx$2 ms and the propagation velocity is $\approx$20 m/s)\cite{Young1936, Hodgkin1952}). Because the characteristics of APs over distances of the order of the pulse size were scarcely investigated, it may be worth to examine these properties in a future work.

\paragraph{Voltage signal of acoustic pulses.} We now turn to examine the effect of density changes on a transmembrane voltage measurement. Changes in electric properties, such as capacitance and resistance, are unavoidable during propagation of acoustic pulses in soft materials\cite{Heimburg2012, Griesbauer2012a}. Furthermore, the surface potential of a lipid monolayer differs by $\sim$100 mV between the liquid-expanded and the liquid-condensed phases, the same order of magnitude as the electric aspect of an AP\cite{Steppich2010}. To quantify these changes, we explore a simple extension to the model; a standard electrophysiological measurement across a lipid interface. For the moment we ignore any transmembrane current of ions, and focus on currents following solely from voltage changes generated by the layer itself. The conservation of charge requires
\begin{equation}
\frac{d}{dt}(\mathbb{C}V)=0,
\end{equation}
 with $\mathbb{C}$ the local capacitance of the material, and $V$ the local transmembrane voltage. The equation implies an inverse relation between the voltage and the material capacitance
 \begin{equation}
\frac{V}{V_i}=\frac{\mathbb{C}_i}{\mathbb{C}},
\end{equation}
with $V_i, \mathbb{C}_i$ the initial voltage and capacitance, respectively. The capacitance of the system can be associated with the area, thickness and relative permittivity of the material ($A, d$ and $\varepsilon$, respectively)
 \begin{equation}
\mathbb{C}\approx\varepsilon_0\varepsilon\frac{A}{d},
\end{equation}
where all variables (including $\varepsilon$) depend on the local thermodynamic state of the membrane\cite{Heimburg2012}. Replacing the area term with the density of the membrane, we obtain a linear relation between the voltage and the local density 
 \begin{equation}
\mathbb{C}\propto\frac{\varepsilon}{d\rho} \Longrightarrow V\propto\frac{d\rho}{\varepsilon}.
\end{equation}
Thus, a propagating density pulse (Fig.~\ref{Fig-01}), if detected by a voltage sensor, should display a distinct shape that includes a depolarization, repolarization and hyperpolarization phase, similar to measurements of APs. However, a voltage pulse is expected to display a sharper non-linear response (a threshold) as compared with the density pulse (Fig.~\ref{Fig-04}b). This is because the relative permittivity and thickness are also state dependent, the former decreases and the latter increases as the pulse traverses into the condensed phase\cite{Steppich2010, Heimburg1998, Kimura1985}. On top of these results, one could include the resistance of the soft system to transmembrane ionic currents, which is governed by a state dependent permeability\cite{El-Mashak1985, Wunderlich2009}. In conclusion, the electric potential difference across a lipid interface responds non-linearly when acoustic pulses travel along the interface. 

The realization that APs may be acoustic pulses, that carry electrical changes as they travel, should have significant implications for the field of neuroscience. For decades the field has placed much focus on measurements and analysis of electric signals in nervous-systems of many organisms. However, it is possible that crucial computational information goes unnoticed when the state of a neuron is simplified into a {\it raster plot} data (a digital-like uniformity of zeros and ones that represent events of APs). Let us describe one plausible scenario. In Fig.~\ref{Fig-04} we have demonstrated that at large excitation amplitudes very similar density pulses can be excited with a significantly different pressure signature. Thus, a voltage sensor, sensitive to density but not to pressure, would not resolve between different pressure signatures that could induce different cellular responses, and result in a considerably different computational scheme as compared to the classical electrical picture. 

\paragraph{Model extensions.} The evidence that the reported pulses do not emerge from a fine-tuned model, but rather result from an idealized description of a soft system, emphasizes the generality of the phenomenon. This, however, should only be viewed as a first step. To treat pulses in soft systems more accurately, further extensions to the model are required. (1) The parameters of the system; specifically, viscosity, heat capacity and thermal conductivity, are state dependent and not constants\cite{Heimburg2005, Hermans2014, Youssefian2017}. This is particularly evident near a phase transition and should modify the detailed properties of the nonlinear pulses. (2) The van der Waals constitutive equation is not an accurate representation of the state diagram of soft materials, lipids in particular. For example, in DPPC the volume exclusion appears at $\approx$0.8$w_c$\cite{Albrecht1978, Heimburg2007, Steppich2010}. This is quantitatively different from the van der Waals equation, where the volume exclusion occurs at $w_c/3$. These differences clearly influence the properties of the pulse. For example, the saturation amplitude of the density pulse would be $\rho_{sat}\approx$1.2$\rho_c$ instead of $\approx$3$\rho_c$. (3) The viscoelastic properties of lipids are more complicated than simply considering a capillarity term\cite{espinosa2011shear}. (4) No effect of geometry was considered in this work (boundary conditions, extension to two- or three-dimensions, coupling to bulk). (5) The parameters that describe the material (critical point, thermal conductivity, etc.)~are very likely influenced by additional components that were not considered here. For example, pH, ions, as well as embedded proteins. (6) In order to provide a detailed analysis of electrophysiological measurements, a more accurate description of the electro-mechanical coupling is necessary. For example, we did not address space- and voltage-clamp experiments in this paper\cite{Hodgkin1952}. (7) Our description is based on a mean field approximation. Therefore, thermodynamic fluctuations were not accounted. In particular, the model does not describe current fluctuations that were experimentally measured in non-living soft systems as well as in living cells\cite{Wunderlich2009}. To accommodate for state fluctuations, an analysis of the thermodynamic susceptibilities is necessary. These aspects should be considered in a future work.

% ----------------- A c k n o w l e d g e m e n t s  ----------------------------
\section{Acknowledgements}
The authors thank Christian Fillafer, Kevin Kang, Julian Kappler and Konrad Kaufmann for fruitful discussions and valuable comments, and to Daniel Lecoanet for support with the Dedalus open-source code. MM thanks Uri Nevo for introducing him to the subject of non-electric aspects of action potentials, and to Jay Fineberg for useful comments. MM further acknowledges funds from SHENC-research unit FOR 1543.

\end{multicols}

% ----------------- R e f e r e n c e s  ----------------------------

\rule{\textwidth}{0.5pt}

\begin{multicols}{2}
\end{multicols}

% ------------- S u p p l e m e n t a l    M a t e r i a l s --------------

\rule{\textwidth}{0.5pt}

\begin{multicols}{2}

\beginsupplement

\begin{appendices}

The supplemental materials are intended to provide interested readers with further mathematical details (section A), additional solutions (section B) and caption to the supplemental movie (section C).

{
\renewcommand{\arraystretch}{1.5}
\begin{table*}
\centering
\begin{tabular}{ | c | c | p{3.5cm} | }
\hline
Variable & Name & Units for a 3d material (2d material) in MKS system \\ \hline
$w$ & specific volume & $\frac{m^3}{kg}~\left(\frac{m^2}{kg}\right)$ \\ \hline
$v$ & velocity & $\frac{m}{s}$ \\ \hline
$E$ & specific total energy & $\frac{m^2}{s^2}$ \\ \hline
$p$ & pressure & $\frac{N}{m^2}~\left(\frac{N}{m}\right)$ \\ \hline
$\theta$ & temperature & $K$ \\ \hline
$\rho$ & density & $\frac{kg}{m^3}~\left(\frac{kg}{m^2}\right)$ \\ \hline
 \end{tabular}
 \caption{Model variables}
 \label{tab:variables}
 \end{table*}
}

{
\renewcommand{\arraystretch}{1.5}
\begin{table*}
\centering
\begin{tabular}{ | c | c | p{3.5cm} | }
\hline
Parameter & Name & Units for a 3d material (2d material) in MKS system \\ \hline
$m$ & mass of a fluid particle & $kg$ \\ \hline
$a$ & average attraction between fluid particles & $Nm^4 \left(Nm^3\right)$ \\ \hline
$b$ & volume exclusion by the particles & $m^3 \left(m^2\right)$ \\ \hline
$\zeta$ & bulk viscosity & $\frac{kg}{m\cdot s}~\left(\frac{kg}{s}\right)$ \\ \hline
$c_v$ & specific heat capacity at constant volume & $\frac{J}{kg\cdot K}$ \\ \hline
$k$ & thermal conductivity & $\frac{J}{m\cdot s\cdot K}~\left(\frac{J}{s\cdot K}\right)$ \\ \hline
$C$ & capillarity coefficient & $\frac{kg^2}{m^2\cdot s^2}~\left(\frac{kg^2}{s^2}\right)$ \\ \hline
 \end{tabular}
 \caption{Model constant parameters}
 \label{tab:parameters}
 \end{table*}
}

% ---------------------- M o d e l    e q u a t i o n s ---------------------------
\section{Model Equations}
\paragraph{Model variables and parameters.} The fluid considered in this letter is described by six dynamic variables of space and time, that are coupled to one another by three conservation laws (Eq.~(1)), two constitutive equations (Eq.~(4)), and one inverse relation, $\rho=w^{-1}$. The six variables are listed in Table \ref{tab:variables}. In addition, the model depends on seven constant parameters, all are measurable physical quantities (Table \ref{tab:parameters}). The van der Waals parameters ($m, a, b$) are related to the critical point according to
\begin{equation}
w_c=\frac{3b}{m}, \qquad p_c=\frac{a}{27b^2}, \qquad \theta_c=\frac{8a}{27b k_B}.
\end{equation}
The scaling of the spatial coordinate $h$ (in the Lagrange frame) was defined according to the critical density ($\bar{w}=w_c$). Finally, $k_B\cong 1.38\cdot 10^{-23}~J/K$ is the Boltzmann constant.
 \begin{figure*}[htb]
\centering
\includegraphics[width=0.5\linewidth]{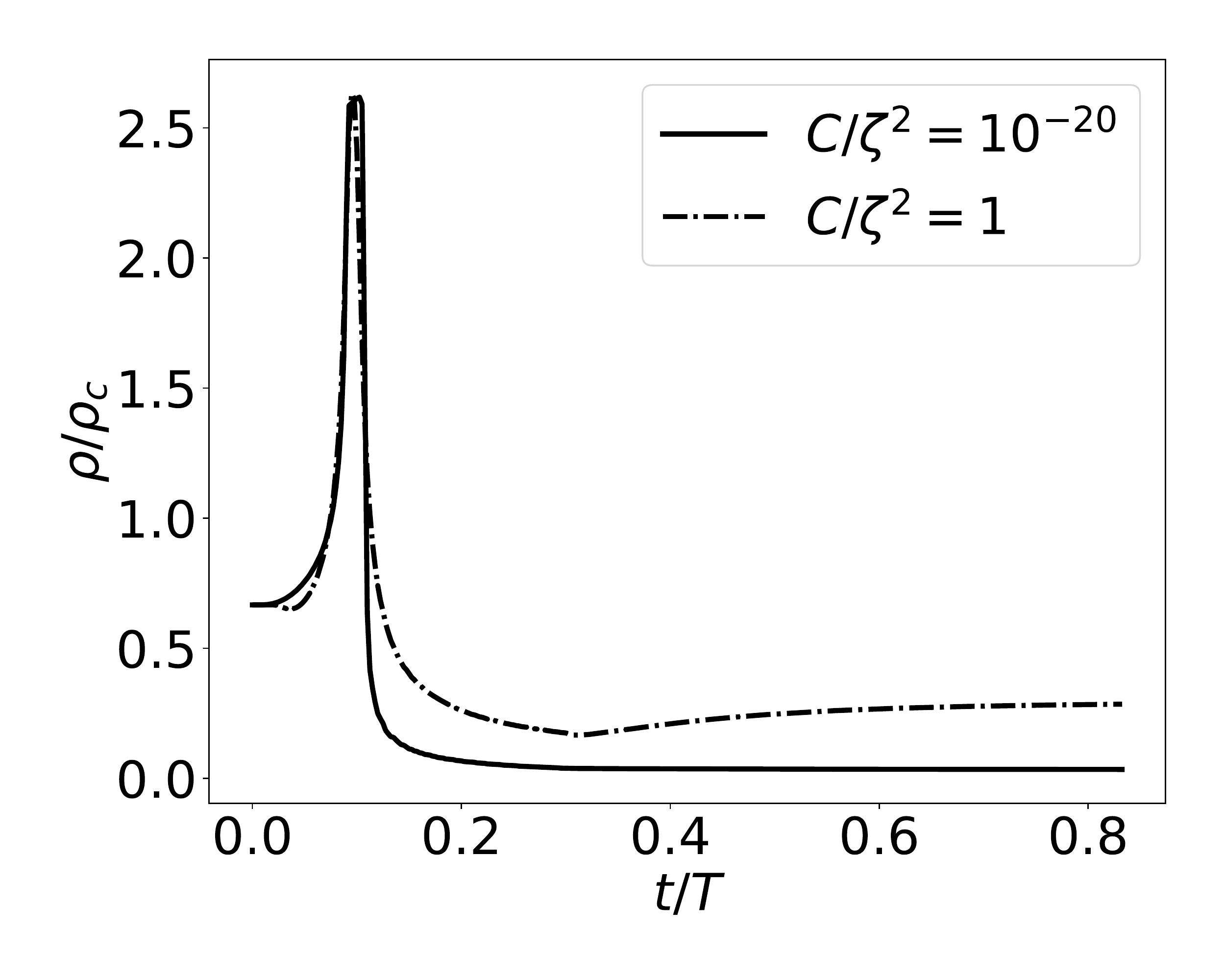}
 \caption{Comparison of pulse shape at two different values of the capillarity coefficient, $C/\zeta^2=10^{-20}$ (solid line) and $1$ (dotted-dashed line). Excitation amplitude was $\tilde{p}_{exc}=60$ and $150$, respectively. Duration of excitation was $\tilde{t}_0=0.3$ for both curves. Pulse was measured at x/L=0.5 and 0.8, respectively. Other parameters appear in caption of Fig.~2.}
 \label{Fig-S1}
 \end{figure*}

\paragraph{Dimensionless model equations.} The critical point and the fluid viscosity parameters were used (Eq.~(5)) to introduce the following reduced variables and parameters:
\begin{equation}\label{eq:dimlessVars}
\begin{aligned}
X\equiv \frac{h}{L},& \qquad \tilde{t}\equiv \frac{t}{T} \\
\tilde{w}\equiv\frac{w}{w_c}, \qquad & \tilde{v}\equiv \frac{v}{U}, \qquad \tilde{E}\equiv\frac{E}{U^2},   \\
\tilde{\theta}\equiv\frac{\theta}{\theta_c}, \qquad &\tilde{p}\equiv\frac{p}{p_c}, \qquad \tilde{\tau}\equiv\frac{\tau}{p_c}.
\end{aligned}
\end{equation}
The spatial dimension in the Lagrange frame was normalized by the choice $\bar{w}=w_c$. The dimensionless model equations are
\begin{equation}\label{eq:dimlessEqs}
\begin{aligned}
\partial_{\tilde{t}}  \tilde{w}&=\partial_X \tilde{v}, \\
\partial_{\tilde{t}}  \tilde{v}&=\partial_X \left(\tilde{\tau}_1+\tilde{\tau}_2 \right),  \\
\partial_{\tilde{t}}  \tilde{E}&=\partial_X \left(\tilde{\tau}_1 \tilde{v} \right)+\tilde{k}\partial_X^2 \tilde{\theta},
\end{aligned}
\end{equation}
with
\begin{equation}\label{eq:stress}
\begin{aligned}
\tilde{\tau}_1 &=-\tilde{p} + \partial_X \tilde{v}, \\
\tilde{\tau}_2 &= -\tilde{C}\partial_X^2\tilde{w},
\end{aligned}
\end{equation}
and
\begin{equation}\label{eq:dimlessvdW}
\begin{aligned}
\tilde{p}&=\frac{8\tilde{\theta}}{3\tilde{w}-1}-\frac{3}{\tilde{w}^2}, \\
\tilde{E}&=\tilde{c}_v \tilde{\theta}-\frac{3}{\tilde{w}} +\frac{\tilde{v}^2}{2}.
\end{aligned}
\end{equation}
The model equations (\ref{eq:dimlessEqs})--(\ref{eq:dimlessvdW}) depend on three dimensionless parameters: heat capacity, thermal conductivity and capillarity coefficient
\begin{equation}
\tilde{c}_v=\frac{c_v \theta_c}{p_c w_c}, \qquad \tilde{k}=\frac{k\theta_c}{p_c w_c \zeta}, \qquad \tilde{C}=\frac{C}{\zeta^2}.
\end{equation}
Note that a physically meaningful solution requires $1/3 < \tilde{w}$ and $0 < \tilde{\theta}$.

\paragraph{Model equations in the Euler frame.} The model equations in the Lagrange frame are given in Eq.~(1). These equations were transformed into the Euler frame using Eq.~(2); i.e., $\partial x/\partial h = w/\bar{w}$, and by replacing the time derivative with the material derivative $\partial_t \rightarrow D_t=\partial_t + v\partial_x$
\begin{equation}
\begin{aligned}
D_t w&=w\partial_x v, \\
D_t v&=w\partial_x (\tau_1+\tau_2 ), \\
D_t E&=w\left[\partial_x (\tau_1 v)+\frac{k}{\bar{w}^2}  w\partial_x (w\partial_x \theta)\right],
\end{aligned}
\end{equation}
and
\begin{equation}
\begin{aligned}
\tau_1&=-p+\zeta\frac{w}{\bar{w}} \partial_x v, \\
\tau_2&=-C\frac{w}{\bar{w}^2} \partial_x (w\partial_x w).
\end{aligned}
\end{equation}
In comparison, the classical Navier-Stokes equations in the Euler frame are
\begin{equation}
\begin{aligned}
D_t w&=w\partial_x v, \\
D_t v&=w\partial_x \tau_{NS}, \\
D_t E&=w[\partial_x (\tau_{NS} v)+k\partial_x^2 \theta],
\end{aligned}
\end{equation}
with
\begin{equation}
\tau_{NS}=-p+\zeta\partial_x v.
\end{equation}

% ---------------------- v d W   f l u i d    e x a m p l e s ---------------------------
 \begin{figure*}[htb]
\centering
\includegraphics[width=0.9\linewidth]{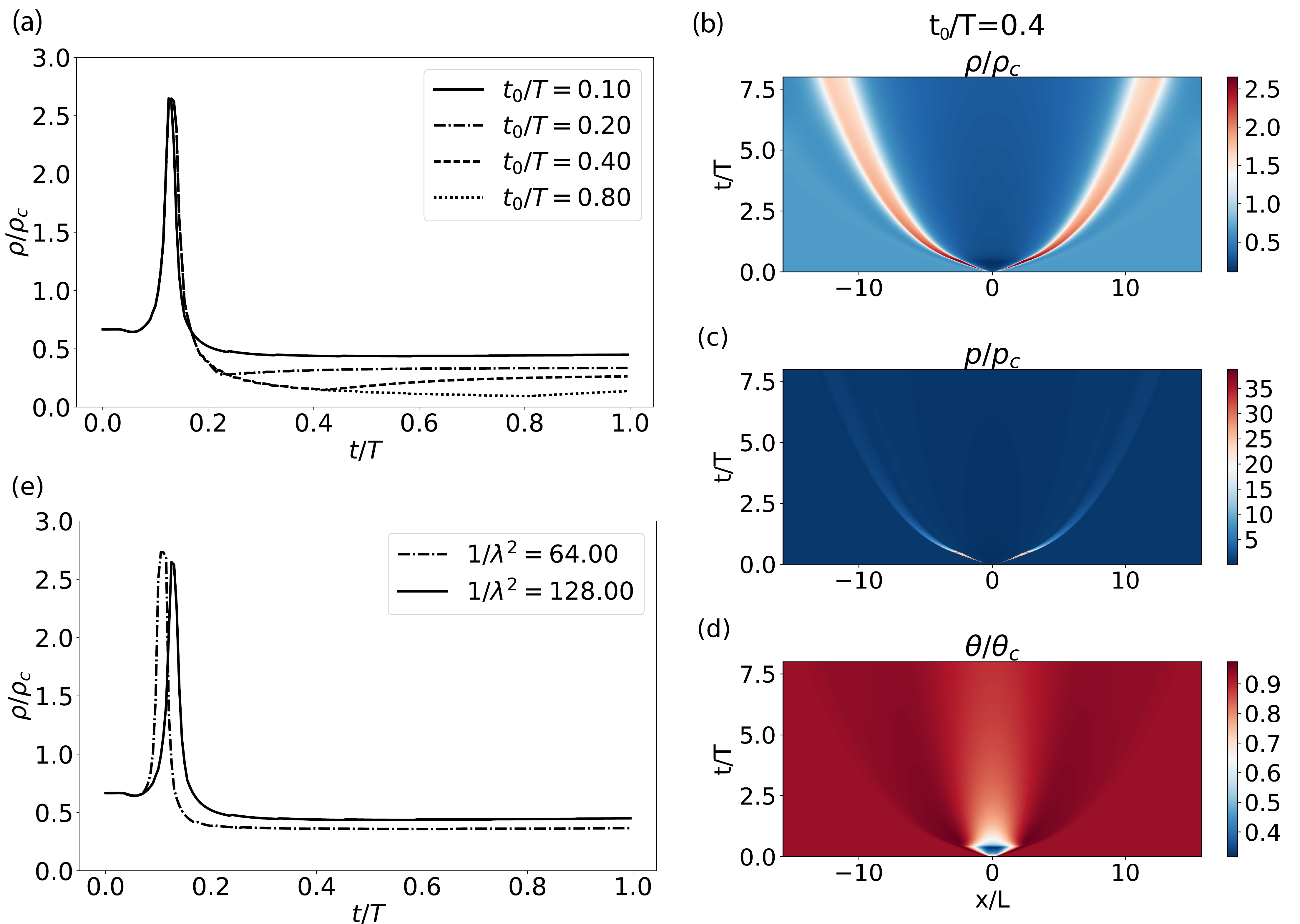}
 \caption{(a) Variation of stimulation duration ($t_0$) did not affect the duration of the compression part of the density pulse, as measured at distance x/L=1 from the excitation point. The subsequent rarefaction, however, was altered. (b) Density, (c) pressure and (d) temperature solutions as a function of space (x-axis) and time (y-axis) upon stimulation with $t_0$/T=0.4. Pulse maintains a recognizable shape for distances larger than 10L. (e) The shape of the pulse did not change much with an increase in stimulation width. Initial density was $\tilde{\rho}_0=0.67$, and numerical calculation was conducted with 2048 grid points, x-domain $[-5\pi, 5\pi]$ and dt=$10^{-3}$. Other parameters appear in the caption of Fig.~2.}
 \label{Fig-S2}
 \end{figure*}
 
\section{Pulse solution in a vdW fluid model}
\paragraph{Capillarity coefficient and the line tension.} Gibbs approached a phase boundary as a infinitesimal line that separates the two phases. The energy stored in a phase boundary of length $L$ and line tension $\gamma$ is $E=\gamma L$. In contrast, van der Waals approached a phase transition as a continuous change, and the phase boundary, therefore, has a finite length. Van der Waals suggested an energy term that depends on the density gradient
 \begin{equation}
 E=\int K\left(\frac{\partial\rho}{\partial x}\right)^2 d^2 x,
 \end{equation}
 with $K\sim C/\rho_c^3$. For a phase boundary of width $\ell$, the energy stored in the phase boundary is $E \sim K \frac{\Delta\rho^2}{\ell^2}\ell L$. Therefore, the capillarity coefficient is related to the line tension according to  
 \begin{equation}
C \sim \frac{\gamma\ell \rho_c^3}{\Delta\rho^2}.
 \end{equation}
 Assuming $\ell\sim10^{-9}-10^{-6}$ m, $\Delta\rho\sim\rho_c\sim 10^{-6}~kg/m^2$ and $\gamma\sim 10^{-12}$ N (Benvegnu and McConnell, J Phys Chem, 1992) we obtain $C\sim 10^{-27}-10^{-24}~kg^2/s^2$. However, in the numerical calculation we have used a larger value ($C/\zeta^2\sim1$), to avoid sharp gradients in density. This did not have qualitative effect on the properties of sound pulses near phase transition as evident in Fig.~\ref{Fig-S1}.
 
 \paragraph{Effect of stimulation parameters on the pulse characteristics.} Figure \ref{Fig-S2}a shows the effect of stimulation duration on the shape of the density pulse, as measured at distance x/L=1. It was evident that the compression part of the density pulse was not affected much and maintained a pulse duration of $\approx$0.1T. On the other hand, the subsequent rarefaction monotonically decreased its density with stimulation duration. The density, pressure and temperature field solutions as a function of space (x-axis) and time (y-axis) are shown in Figs~\ref{Fig-S2}b,c,d, respectively, for stimulation duration of $t_0$/T=0.4. It was evident that the pulse maintained a recognizable shape for distances larger than 10L. An increase of the stimulation width ($\lambda$) resulted in similar observations (Fig.~\ref{Fig-S2}e). 
  
% ---------------------- M o v i e    C a p t i o n s --------------------------- 
\section{Caption of supplemental movie}
\underline{Movie S1:} Collision between two propagating pulses is demonstrated in the density, pressure and temperature fields. At the peak of the collision a short burst of pressure is obtained (t/T=0.2). The pulses annihilate afterwards. Computational details are given in caption of Fig.~5. 

\end{appendices}

\end{multicols}


{
\footnotesize
\begin{thebibliography}{10}

\bibitem{Tasaki1982}
I.~Tasaki, {\em {Physiology and electrochemistry of nerve fibers}}.
\newblock New York: Academic press, 1982.

\bibitem{Mackie2004}
G.~O. Mackie, ``{Epithelial conduction: Recent findings, old questions, and
  where do we go from here?},'' {\em Hydrobiologia}, vol.~530-531, pp.~73--80,
  2004.

\bibitem{Parak1999}
W.~J. Parak, J.~Domke, M.~George, a.~Kardinal, M.~Radmacher, H.~E. Gaub, a.~D.
  de~Roos, a.~P. Theuvenet, G.~Wiegand, E.~Sackmann, and J.~C. Behrends,
  ``{Electrically excitable normal rat kidney fibroblasts: A new model system
  for cell-semiconductor hybrids.},'' {\em Biophysical journal}, vol.~76,
  no.~3, pp.~1659--1667, 1999.

\bibitem{fields2008oligodendrocytes}
R.~D. Fields, ``{Oligodendrocytes changing the rules: action potentials in glia
  and oligodendrocytes controlling action potentials},'' {\em The
  Neuroscientist}, vol.~14, no.~6, pp.~540--543, 2008.

\bibitem{Ashcroft1989}
F.~M. Ashcroft and P.~Rorsman, ``{Electrophysiology of the pancreatic
  $\beta$-cell},'' {\em Progress in Biophysics and Molecular Biology}, vol.~54,
  no.~2, pp.~87--143, 1989.

\bibitem{Leys2015}
S.~P. Leys, ``{Elements of a 'nervous system' in sponges.},'' {\em The Journal
  of experimental biology}, vol.~218, no.~Pt 4, pp.~581--91, 2015.

\bibitem{Beilby2007}
M.~J. Beilby, ``{Action potential in charophytes},'' {\em International review
  of cytology}, vol.~257, pp.~43--82, jan 2007.

\bibitem{slayman1976action}
C.~L. Slayman, W.~S. Long, and D.~Gradmann, ``{?Action potentials? in
  Neurospora crassa, a mycelial fungus},'' {\em Biochimica et Biophysica Acta
  (BBA)-Biomembranes}, vol.~426, no.~4, pp.~732--744, 1976.

\bibitem{Wood1982}
D.~C. Wood, ``{Membrane permeabilities determining resting, action and
  mechanoreceptor potentials in Stentor coeruleus},'' {\em Journal of
  Comparative Physiology}, vol.~146, no.~4, pp.~537--550, 1982.

\bibitem{kralj2011electrical}
J.~M. Kralj, D.~R. Hochbaum, A.~D. Douglass, and A.~E. Cohen, ``{Electrical
  spiking in Escherichia coli probed with a fluorescent voltage-indicating
  protein},'' {\em Science}, vol.~333, no.~6040, pp.~345--348, 2011.

\bibitem{Hodgkin1952}
A.~L. Hodgkin and A.~F. Huxley, ``{A Quantitative Description of Membrane
  Current and Its Application to Conduction and Excitation in Nerve},'' {\em
  Journal of Physiology}, vol.~117, pp.~500--544, 1952.

\bibitem{aidley1998physiology}
D.~J. Aidley and D.~J. Ashley, {\em {The physiology of excitable cells}},
  vol.~4.
\newblock Cambridge University Press Cambridge, 1998.

\bibitem{Shrivastava2014}
S.~Shrivastava and M.~F. Schneider, ``{Evidence for 2D Solitary Sound Waves in
  a Lipid Controlled Interface and its Biological Implications for Biological
  Signaling},'' {\em Journal of The Royal Society Interface}, vol.~11,
  pp.~1--8, 2014.

\bibitem{Przybylski1982}
A.~T. Przybylski, W.~P. Stratten, R.~M. Syren, and S.~W. Fox, ``{Membrane,
  action, and oscillatory potentials in simulated protocells},'' {\em
  Naturwissenschaften}, vol.~69, no.~12, pp.~561--563, 1982.

\bibitem{Tasaki1999}
I.~Tasaki, ``{Rapid Structural Changes in Nerve Fibers and Cells Associated
  With Their Excitation Processes},'' {\em Japanese Journal of Physiology},
  vol.~49, pp.~125--138, 1999.

\bibitem{Tasaki1966}
I.~Tasaki, A.~Watanabe, and I.~Singer, ``{Excitability of Squid Giant Axon in
  the Absence of Univalent Cations in the External Medium},'' {\em Proceedings
  of the National Academy of Sciences}, vol.~56, pp.~1116--1122, 1966.

\bibitem{Terakawa1981}
S.~Terakawa, ``{Periodic responses in squid axon membrane exposed
  intracellularly and extracellularly to solutions containing a single species
  of salt},'' {\em The Journal of Membrane Biology}, vol.~63, no.~1-2,
  pp.~51--59, 1981.

\bibitem{Wunderlich2009}
B.~Wunderlich, C.~Leirer, A.~L. Idzko, U.~F. Keyser, A.~Wixforth, V.~M. Myles,
  T.~Heimburg, and M.~F. Schneider, ``{Phase-state dependent current
  fluctuations in pure lipid membranes},'' {\em Biophysical Journal}, vol.~96,
  no.~11, pp.~4592--4597, 2009.

\bibitem{Heimburg2005}
T.~Heimburg and A.~D. Jackson, ``{On Soliton Propagation in Biomembranes and
  Nerves},'' {\em Proceedings of the National Academy of Sciences}, vol.~102,
  no.~28, pp.~9790--9795, 2005.

\bibitem{Kaufmann1989}
K.~Kaufmann, ``{Action Potentials and Electromechanical Coupling in the
  Macroscopic Chiral Phospholipid Bilayer},'' 1989.

\bibitem{ling1992revolution}
G.~N. Ling, ``{A Revolution in the Physiology of the Living Cell},'' 1992.

\bibitem{Hodgkin1945}
A.~L. Hodgkin and A.~F. Huxley, ``{Resting and Action Potentials in Single
  Nerve Fibers},'' {\em J. Physiol.}, vol.~104, pp.~176--195, 1945.

\bibitem{Wikswo1980}
J.~P. Wikswo, J.~P. Barach, and J.~A. Freeman, ``{Magnetic field of a nerve
  impulse: first measurements.},'' {\em Science (New York, N.Y.)}, vol.~208,
  no.~4439, pp.~53--55, 1980.

\bibitem{Steppich2010}
D.~Steppich, J.~Griesbauer, T.~Frommelt, W.~Appelt, A.~Wixforth, and M.~F.
  Schneider, ``{Thermomechanic-electrical coupling in phospholipid monolayers
  near the critical point.},'' {\em Physical Review E}, vol.~81, no.~6 Pt 1,
  p.~061123, 2010.

\bibitem{Griesbauer2012}
J.~Griesbauer, S.~B{\"{o}}ssinger, A.~Wixforth, and M.~F. Schneider,
  ``{Propagation of 2D Pressure Pulses in Lipid Monolayers and Its Possible
  Implications for Biology},'' {\em Physical Review Letters}, vol.~108, no.~19,
  p.~198103, 2012.

\bibitem{Shrivastava2015}
S.~Shrivastava, K.~H. Kang, and M.~F. Schneider, ``{Solitary shock waves and
  adiabatic phase transition in lipid interfaces and nerves},'' {\em Physical
  Review E}, vol.~91, no.~1, pp.~1--7, 2015.

\bibitem{Fichtl2016}
B.~Fichtl, S.~Shrivastava, and M.~Schneider, ``{Protons at the speed of sound:
  Predicting specific biological signaling from physics},'' {\em Scientific
  Reports}, vol.~6, no.~1, p.~22874, 2016.

\bibitem{Ritchie1985}
J.~M. Ritchie and R.~D. Keynes, ``{The production and absorption of heat
  associated with electrical activity in nerve and electric organ},'' {\em
  Quarterly Reviews of Biophysics}, vol.~18, no.~4, pp.~451--476, 1985.

\bibitem{Heimburg2007}
T.~Heimburg, {\em {Thermal Biophysics of Membranes}}.
\newblock Wiley-VCH, 1st~ed., 2007.

\bibitem{Griesbauer2009}
J.~Griesbauer, A.~Wixforth, and M.~F. Schneider, ``{Wave propagation in lipid
  monolayers},'' {\em Biophysical Journal}, vol.~97, no.~10, pp.~2710--2716,
  2009.

\bibitem{Griesbauer2012a}
J.~Griesbauer, S.~B{\"{o}}ssinger, A.~Wixforth, and M.~F. Schneider,
  ``{Simultaneously Propagating Voltage and Pressure Pulses in Lipid Monolayers
  of Pork Brain and Synthetic Lipids},'' {\em Physical Review E}, vol.~86,
  no.~6, pp.~1--5, 2012.

\bibitem{Hodgkin1952a}
A.~L. Hodgkin, A.~F. Huxley, and B.~Katz, ``{Measurement of current-voltage
  relations in the membrane of the giant axon of Loligo},'' {\em The Journal of
  Physiology}, vol.~116, no.~4, pp.~424--448, 1952.

\bibitem{Tasaki1949}
I.~Tasaki, ``{Collision of two nerve impulses in the nerve fibre},'' {\em
  Biochimica et Biophysica Acta}, vol.~3, pp.~494--497, 1949.

\bibitem{Follmann2015}
R.~Follmann, E.~Rosa, and W.~Stein, ``{Dynamics of signal propagation and
  collision in axons},'' {\em Physical Review E}, vol.~92, no.~3, pp.~1--11,
  2015.

\bibitem{Shrivastava2017}
S.~Shrivastava, K.~H. Kang, and M.~F. Schneider, ``{Collision and Annihilation
  of Nonlinear Pulses and Action Potentials in Interfaces},'' {\em
  unpublished}, 2017.

\bibitem{Kappler2017}
J.~Kappler, S.~Shrivastava, M.~F. Schneider, and R.~R. Netz, ``{Nonlinear
  fractional waves at elastic interfaces},'' {\em arXiv: 1702.08864}, 2017.

\bibitem{slemrod1984dynamic}
M.~Slemrod, ``{Dynamic phase transitions in a van der Waals fluid},'' {\em
  Journal of differential equations}, vol.~52, no.~1, pp.~1--23, 1984.

\bibitem{grinfeld1989nonisothermal}
M.~Grinfeld, ``{Nonisothermal dynamic phase transitions},'' {\em Quarterly of
  Applied Mathematics}, vol.~47, no.~1, pp.~71--84, 1989.

\bibitem{felderhof1970dynamics}
B.~U. Felderhof, ``{Dynamics of the diffuse gas-liquid interface near the
  critical point},'' {\em Physica}, vol.~48, no.~4, pp.~541--560, 1970.

\bibitem{Albrecht1978}
O.~Albrecht, H.~Gruler, and E.~Sackmann, ``{Polymorphism of phospholipid
  monolayers},'' {\em Journal de Physique}, vol.~39, no.~3, pp.~301----313,
  1978.

\bibitem{courant1948supersonic}
R.~Courant and K.~O. Friedrichs, {\em {Supersonic flow and shock waves}},
  vol.~21.
\newblock Wiley-Interscience, New York, 1948.

\bibitem{Gorban2016}
A.~N. Gorban and I.~V. Karlin, ``{Beyond Navier?Stokes equations: capillarity
  of ideal gas},'' {\em Contemporary Physics}, vol.~7514, no.~November,
  pp.~1--21, 2016.

\bibitem{van20thermodynamic}
J.~D. van~der Waals, ``{The thermodynamic theory of capillarity flow under the
  hypothesis of a continuous variation of density (Verhandel/Konink. Akad.
  Weten., 1893, vol. 1, English Translation)},'' {\em Journal of Statistical
  Physics}, vol.~20.

\bibitem{landau1980statistical}
L.~D. Landau and E.~M. Lifshitz, ``{Statistical physics, vol. 5},'' {\em Course
  of theoretical physics}, vol.~30, 1980.

\bibitem{cahn1958free}
J.~W. Cahn and J.~E. Hilliard, ``{Free energy of a nonuniform system. I.
  Interfacial free energy},'' {\em The Journal of chemical physics}, vol.~28,
  no.~2, pp.~258--267, 1958.

\bibitem{korteweg1901forme}
D.~J. Korteweg, ``{Sur la forme que prennent les {\'{e}}quations du mouvement
  des fluides si l'on tient compte des forces capillaires caus{\'{e}}es par des
  variations de densit{\'{e}} consid{\'{e}}rables mais continues et sur la
  th{\'{e}}orie de la capillarit{\'{e}} dans l'hypothese d'une variatio},''
  {\em Archives N{\'{e}}erlandaises des Sciences exactes et naturelles},
  vol.~6, no.~1, p.~6, 1901.

\bibitem{johnston2014thermodynamic}
D.~C. Johnston, ``{Thermodynamic properties of the van der Waals fluid},'' {\em
  arXiv:1402.1205}, 2014.

\bibitem{espinosa2011shear}
G.~Espinosa, I.~L{\'{o}}pez-Montero, F.~Monroy, and D.~Langevin, ``{Shear
  rheology of lipid monolayers and insights on membrane fluidity},'' {\em
  Proceedings of the National Academy of Sciences}, vol.~108, no.~15,
  pp.~6008--6013, 2011.

\bibitem{clegg1991biochemistry}
J.~S. Clegg and W.~Drost-Hansen, ``{On the biochemistry and cell physiology of
  water-Chapter 1},'' 1991.

\bibitem{Benvegnu1992}
D.~J. Benvegnu and H.~M. Mcconnell, ``{Line Tension between Lipid Domains in
  Lipid Monolayers},'' {\em Journal of Physical Chemistry}, vol.~96,
  pp.~6820--6824, 1992.

\bibitem{Burns2017}
K.~J. Burns, G.~M. Vasil, J.~S. Oishi, D.~Lecoanet, B.~P. Brown, and
  E.~Quataert, ``{Dedalus: A Flexible Framework for Spectrally Solving
  Differential Equations},'' {\em (unpublished)}, 2017.

\bibitem{Bakkum2013}
D.~J. Bakkum, U.~Frey, M.~Radivojevic, T.~L. Russell, J.~M{\"{u}}ller,
  M.~Fiscella, H.~Takahashi, and A.~Hierlemann, ``{Tracking axonal action
  potential propagation on a high-density microelectrode array across hundreds
  of sites.},'' {\em Nature communications}, vol.~4, p.~2181, 2013.

\bibitem{patolsky2006detection}
F.~Patolsky, B.~P. Timko, G.~Yu, Y.~Fang, A.~B. Greytak, G.~Zheng, and C.~M.
  Lieber, ``{Detection, stimulation, and inhibition of neuronal signals with
  high-density nanowire transistor arrays},'' {\em Science}, vol.~313,
  no.~5790, pp.~1100--1104, 2006.

\bibitem{Young1936}
J.~Z. Young, ``{The structure of nerve fibres in Cephalopods and Crustacea},''
  {\em Proceedings of the Royal Society B: Biological Sciences}, vol.~121,
  no.~823, pp.~319--337, 1936.

\bibitem{Heimburg2012}
T.~Heimburg, ``{The capacitance and electromechanical coupling of lipid
  membranes close to transitions: The effect of electrostriction},'' {\em
  Biophysical Journal}, vol.~103, no.~5, pp.~918--929, 2012.

\bibitem{Heimburg1998}
T.~Heimburg, ``{Mechanical aspects of membrane thermodynamics. Estimation of
  the mechanical properties of lipid membranes close to the chain melting
  transition from calorimetry},'' {\em Biochimica et Biophysica Acta -
  Biomembranes}, vol.~1415, no.~1, pp.~147--162, 1998.

\bibitem{Kimura1985}
Y.~Kimura and a.~Ikegami, ``{Local dielectric properties around polar region of
  lipid bilayer membranes.},'' {\em The Journal of membrane biology}, vol.~85,
  no.~3, pp.~225--231, 1985.

\bibitem{El-Mashak1985}
E.~El-Mashak and T.~Tsong, ``{Ion selectivity of temperature - induced and
  electric field induced pore in dipalmitoylphosphatidylcholine vesicles},''
  {\em Biochemistry}, vol.~24, pp.~2884--2888, 1985.

\bibitem{Hermans2014}
E.~Hermans and J.~Vermant, ``{Interfacial shear rheology of DPPC under
  physiologically relevant conditions.},'' {\em Soft matter}, vol.~10,
  pp.~175--186, 2014.

\bibitem{Youssefian2017}
S.~Youssefian, N.~Rahbar, C.~R. Lambert, and S.~V. Dessel, ``{Variation of
  thermal conductivity of DPPC lipid bilayer membranes around the phase
  transition temperature},'' {\em J. R. Soc. Interface}, vol.~14, p.~20170127,
  2017.

\end{thebibliography}
}
\end{document}